\documentclass[prl,aps,amsmath,amssymb,letter]{revtex4}
\usepackage{graphicx}% Include figure files
\usepackage{epsfig,latexsym}
\begin{document}
\title{Formation of Cosmic String network from black holes: Implications
from liquid crystal experiments}
\author{Ajit M. Srivastava}
\email{ajit@iopb.res.in}
\affiliation{Institute of Physics, Sachivalaya Marg, 
Bhubaneswar 751005, India}
% IOP Preprint number: IP/BBSR/2006-10, July 2006.

\begin{abstract}
   We present observation of  large, expanding string loops forming
around a heated wire tip embedded in a nematic liquid crystal sample.
Loops expand due to convective stretching. This observation leads to
a new insight into phenomena which could occur in the early universe.
We show that local heating of plasma in the early universe by evaporating
primordial black holes can lead to formation of large, expanding cosmic
string loops, just as observed in the liquid crystal experiment. 
Intercommutation of string loops from neighboring black holes can 
lead to percolation, thereby forming an infinite string network. This is 
remarkable as such an infinite string network is thought to arise only 
when the entire universe undergoes phase transition.
\end{abstract}

\maketitle

 Condensed matter physics experiments can provide very fruitful
analogues of phenomena thought to have occurred in the early 
universe. An important example of this is provided by the realization 
that the theories of formation of exotic topological defects, such as 
cosmic strings in the early universe \cite{kbl} 
can be tested in condensed matter systems \cite{zrk}. Many experimental
studies have been carried out in liquid crystals \cite{nlc1}, superfluid 
helium \cite{he43}, and superconductors \cite{sc}, by looking 
for analogies of cosmic phenomenon, as well as by focusing on 
universal properties for rigorous quantitative tests of cosmic defect 
theories \cite{nlc2,nlc3}, in  these systems.  
Here, we demonstrate the tremendous
potential of this correspondence by showing that a completely new
insight into the phenomena in the early universe is obtained from
a liquid crystal experiment, namely, the formation of large
string loops due to convective stretching in a liquid crystal sample with
a heated wire tip. We show that the same physics implies that local
heating of plasma in the early universe by evaporating  primordial black
holes \cite{pbh} can lead to formation of large cosmic string loops
which can percolate to form an infinite string network (even when ambient
temperature is far below the cosmic string scale). This is remarkable as
such an infinite string network is thought
to arise only when the entire universe undergoes phase transition.

  Formation of cosmic strings is of great importance as they can provide 
a direct possible window to the physics of ultra-high energy scales.
Recent realization that superstring theories can also lead to existence 
of cosmic strings \cite{superst} has further added to the motivation of 
understanding various mechanisms of formation of cosmic strings. In the 
conventional picture, strings are produced when the universe goes through 
a symmetry breaking phase transition \cite{kbl}. This may not always 
be possible, for example, in models of inflation which have low reheat 
temperature, or in models of TeV scale gravity \cite{tev}.  Our results are 
therefore important as they provide a new possibility where an infinite 
string network can form without  an over all phase transition and when 
the ambient temperature is much lower than the cosmic string scale.

Our experimental setup consists of
a droplet (size $\sim$ 2-3 mm) of uniaxial nematic liquid crystal (NLC) 
$4'$-Pentyl-4-biphenyl-carbonitrile (98\% pure, purchased from Aldrich 
Chem.) placed on a clean, untreated glass slide. For uniaxial nematic
liquid crystals the order parameter is called the director which is a 
vector with opposite directions identified. This arises due to
rod like nature of the molecules in this case (or, if the molecules
are disk like, they are stacked to form a rod like structure). In
the low temperature phase (the nematic phase), molecules tend to
align locally, leading to non zero magnitude of the order parameter
with the director representing the average orientation of the
molecules in a local region. In the high temperature phase (the
isotropic phase), which occurs above a critical value of the temperature,
$T_c$, molecules are randomly oriented leading to vanishing
magnitude of the order parameter. This leads to the order
parameter space being $RP^2 (\equiv S^2/Z_2)$ which allows for
string defects with winding 1/2. The director rotates by $\pi$ around
this string defect.

The isotropic-nematic (I-N) transition temperature $T_c$ for the NLC 
we have used  is about 35.3 $^0$C. The sample was kept at
the ambient temperature less than $T_c$ so it was in the nematic phase.
A copper wire of about 100 micron thickness, with  shape {\Huge $\neg$} 
was connected to a resistor for heating (using voltage supply) such that 
the tip of the wire could be lowered into the sample drop from the top.
The wire tip was in between the objective of the microscope
and the sample. By increasing the current, the wire tip could be
heated to a temperature much higher than $T_c$ (upto about 50 - 60
$^0$ C). A Leica DMRM microscope with 20x objective, and a CCD camera
was used for observations, at the Institute of Physics, Bhubaneswar. The
string formation was recorded using a camcorder (for larger field of view).

When the current is increased with the wire tip inside
the NLC drop then a small region around the tip heats up above $T_c$
to the isotropic phase. When the current is low such that this isotropic 
region is small then small string loops form and shrink near the wire tip,
as seen in Fig.1a.
As convection currents take the liquid (crystal) elements away
from the heated tip, the temperature of the liquid element drops and falls
below $T_c$ at a certain distance from the tip. Resulting I-N phase 
transition can lead to string formation via the standard  Kibble mechanism 
\cite{kbl}. Another way strings can form here is because of turbulence
generated by the heated tip due to temperature gradients in both
vertical and horizontal directions. This can lead to formation of string 
defects as the order parameter (director) undergoes
non-trivial rotations due to turbulent vorticity.

 Fig. 1b-d  show large string loops further away from the wire tip as the 
wire tip is heated further. (Tip appears bigger in the picture when it 
is embedded deeper in the sample.) Larger heating increases string formation. 
At the same time we see that string loops, instead of shrinking, move away 
from the tip and stretch to large sizes. This is surprising, as one
expects string loops in liquid crystal to always shrink due to string 
tension, (i.e. due to excess free energy). This is because, unlike superfluid 
vortices, there are no velocity fields associated with the winding of 
the liquid crystal string defects. What happens here is that strings 
are swept away from the wire tip by the radial convection currents which 
force strings to stretch and become larger. Eventually string loops
shrinks, presumably when the fluid velocity subsides and is unable to
compensate for the string tension (i.e., the tendency for string length
to shrink due to excess free energy). Due to vertical temperature
gradient, the convection currents also force string loops to return 
towards the wire tip. (Interestingly, confirmation for this pattern of
convection currents is provided by observation of  dirt particles 
embedded in the sample which are seen to follow the motion of string 
loops.) Often we see that string loops get trapped in cyclic 
motions, repeatedly swept away from the tip and becoming larger, then 
shrinking while returning towards the tip. We have checked that when the 
tip is kept just outside the sample (so there are no boundary effects
due to the tip inside the sample) then string loops are still produced,
though lesser in number (which, at least partly is due to lesser heating).

  There is a natural analog of this phenomenon in the early 
universe, where the Hawking evaporation of primordial black holes
\cite{pbh} can lead to local heating of the ambient plasma
\cite{ewb,ewb2} (see, also, \cite{evap}), leading to 
restoration of various symmetries.  An investigation 
of this local heating was carried out in \cite{ewb} utilizing
known results about the energy loss of quarks and gluons traversing 
a region of quark-gluon plasma \cite{eloss}. (We mention here that the
results of ref.\cite{eloss} are certainly applicable for high energy partons 
(say several 100 GeV), it is not clear whether they can be extrapolated to 
the case of Hawking radiation with energies almost near the GUT scale. 
Still, the overall picture of pbh heating the local plasma 
to very high temperatures 
may remain valid even for such large Hawking temperatures).  It was 
shown in \cite{ewb} that this energy loss leads to rapid heating of the
plasma near the black hole. Further, resulting temperature gradients
lead to large pressure gradients such that plasma develops radial flow.

 Possibility of formation of primordial black holes, and various 
observational constraints on them are well discussed in the literature, 
see refs. \cite{pbh} for early discussions.
In first order phase transitions, particularly in first order inflationary
scenarios, primordial black holes can be produced by collapsing regions of
false vacuum \cite{kodamaetal} or due to inhomogeneities formed during
bubble wall collisions \cite{hawkingetal}. They could also
form by large amplitude density
perturbations produced due to fluctuations in the inflaton field 
\cite{carrlidsey}, or by shrinking cosmic string loops
\cite{hawking2}.

 For primordial black holes produced from large amplitude density 
fluctuations produced by the inflation, one might ask  whether CMBR 
data puts any constraints. However, as discussed in ref. \cite{green},
one should note that CMBR data imposes restrictions on the power 
spectrum at {\it large scales} (of order 1 - $10^3$ Mpc), whereas, 
the primordial black holes form due to large amplitude and {\it much
smaller  scale} density fluctuations produced by the inflation. Hence 
they do not affect CMBR results at all but imposes constraints on the 
spectral index and consequently puts restrictions on certain models of 
inflation.

 The constraints on primordial black hole formation come from other sources
like nucleosynthesis, gamma ray background etc. \cite{green}.
However, all these phenomenon essentially imposes constraints on primordial
black holes with a mass range which is much larger than the ones 
relevant here. We consider primordial black holes with masses at most
few tonnes and such black holes evaporate away much above the QCD scale,
without causing any conflict with current observations.
Our main aim is to illustrate a completely novel possibility of forming
infinite string networks via back holes. Hence, we will simply assume
the required masses and number densisties of black holes, without
discussing any specific mechanism which could give rise to the formation 
of such black holes.

Consider the case when the plasma heated by the black hole leads to
local restoration of a gauge symmetry (with an energy scale and associated
phase transition temperature of order $\eta$) which allows for the 
existence of cosmic strings \cite{kbl}.  As the plasma 
flows out radially away from the black hole, its temperature will decrease 
and fall below $T_c (= \eta)$ at some distance $r_\eta$ where the symmetry 
will be broken. Near that region, again, string defects will form either
via the Kibble mechanism, or due to turbulence. (For internal symmetries,
spatial variations of the order parameter could lead to formation of
non-trivial windings when turbulent motion of the plasma folds up
extended spatial regions, essentially compactifying parts of spatial 
regions, i.e. plasma). If the friction forces dominate over the
string tension so that strings are  effectively frozen locally in the
plasma then the string loops will be carried out by the radially expanding 
plasma. They will then stretch out to large sizes instead of shrinking
due to tension, just as for the liquid crystal experiment in Fig.1.
It is important here to realize the difference between string tension for
the liquid crystal example and for the relativistic cosmic string case.
For liquid crystals, string tension represents increased free energy
associated with the string defect configuration. String shrinks here as the 
free energy of the system decreases. For cosmic strings, it is the 
relativistic energy of the cosmic string configuration (arising from
the effective potential) . String tension leads
to shrinking of cosmic string when energy can be dissipated in other modes, 
such as gravitational waves, or dissipation in background plasma. 
The final effect is the same that the strings like to shrink. In our case,
strings expand because plasma flow increases the energy (free energy for
liquid crystals, and relativistic energy for cosmic strings).

If there is a uniform density of similar primordial black holes, then each 
black hole will emit string loops which will be stretched to large sizes 
as they are carried away by convective flow, and may intersect each other. 
It is well established that string defects when crossing each other, 
intercommute \cite{crs} (unless they have ultra-relativistic 
velocities, which is not the case here).  Intercommutation of 
string loops from different black holes will result in formation 
of larger string loops, leading to the possibility that these string 
loops may percolate and form an infinite string network.

  Let us discuss the conditions for percolation of
strings. A black hole of mass $M_{bh}$ evaporates 
by emitting Hawking radiation with an associated
temperature $T_{bh} = {M_{Pl}^2 \over 8 \pi M_{bh}} $ where $M_{Pl} = 1.2
\times 10^{19} GeV$ is the Planck mass. We use natural units with
$\hbar$ = c = 1. The rate of loss of mass by the
evaporating black hole is given by
                                                                                
\begin{equation}
{dM_{bh} \over dt} = - {\alpha M_{Pl}^4 \over M_{bh}^2} ~.
\end{equation}

Here, $\alpha$ accounts for the scattering of emitted particles by
the curvature and depends on $T_{bh}$. For values of $T_{bh}$
relevant for our model, we take $\alpha$ to be $\sim 3 \times 10^{-3}$ 
(see, \cite{ewb,alpha}). The lifetime $\tau_{bh}$ of the black hole 
obtained from Eq.(1) is $\tau_{bh} \simeq 10^2  M_{Pl}^{-4} M_0^3$.
where $M_0$ is the initial mass of the black hole.
Eq.(1) implies that very little energy is emitted until time of the order of
$\tau_{bh}$ which is when  most of the energy gets emitted.
The age of the Universe $t_U$ when its temperature is $T_U$ is \cite{kolb}
$t_U \simeq 0.3 g_*^{-1/2} M_{Pl} T_U^{-2}$, where $g_* (\simeq 100$) is the 
number of relevant degrees of freedom \cite{kolb}. By equating $\tau_{bh} 
= t_U$, we can get the mass $M_0$ of the black hole such that its 
evaporation becomes effective when the temperature of the Universe is $T_U$,

\begin{equation}
M_0 =0.07 M_{Pl}^{5/3} T_U^{-2/3} ~.
\end{equation}
                                                                                
For example, with $T_U$ = 1 GeV we get $M_0 = 4 \times 10^{11} M_{Pl}$, $T_{bh}
= 10^6$ GeV and $\tau_{bh} = 5\times 10^{17} GeV^{-1}=
3 \times 10^{-7}$ s.  (We mention here that the mass of a primordial black 
hole can also increases due to accretion of background plasma particles.
However, for relevant ranges of black hole masses here, all the dominant
particle species  are ultra relativistic, so only the geometric cross-section
of black hole is relevant which is not very effective \cite{acrt}.
Some slow growth of the mass of the black hole could occur because of
this, and the black hole masses we use here should be taken to be the final
mass of the black hole when its evaporation becomes effective.)

In ref. \cite{ewb}, conditions for the equilibration of 
the Hawking radiation in the ambient plasma were discussed. Further, using
the Euler equation for a relativistic fluid, it was shown in \cite{ewb}
that plasma velocity rapidly becomes relativistic due to
large temperature gradients near the black hole.
Assuming a steady state situation where the luminosity $L(r)$ is 
independent of the distance $r$ from black hole center,  and is 
equal to $-dM_{bh}/dt$ (Eq.(1)), the temperature profile $T(r)$ of 
the expanding plasma was obtained in \cite{ewb} to be,

\begin{equation}
T(r) =\Biggl[{L\over (8\pi^3 g_*/45) r^2 v_p }\Biggr]^{1\over4} ~.
\end{equation}

This was valid for distances where bulk plasma flow dominates over 
diffusion of particles \cite{ewb}, and will be applicable for 
distances relevant here. We will take the plasma velocity 
$v_p$ to be of order of sound velocity $ v_p \simeq 1/\sqrt{3}$. We are 
neglecting shock formation here for simplicity, though shocks may be more 
favorable for string percolation. 

 With Eq.(1) we see that it is the last stages
of black hole evaporation which will lead to highest temperatures
for the plasma in the nearby region, and will be most relevant for
our model.  Thus, we consider the situation when a black hole of initial 
mass $M_0$ (Eq.(2)) has reduced by evaporation to a much
smaller black hole with a mass  $M_x \equiv M_0/x$, whose life time is
$\tau_x$ and temperature is $T_x > \eta$. Here $x > 1$, and is constrained 
by requiring that the thickness of the heated plasma region (which is of 
the order of $\tau_x$ for plasma expanding with sound velocity)
should be greater than $\eta^{-1}$ (for self consistency of string 
formation at scale $\eta$).  We thus require.

\begin{equation}
100 M_{pl}^{-4} {M_0^3 \over x^3} \ge \eta^{-1} ~.
\end{equation}

 Temperature profile of the plasma around this remaining black hole is obtained
from Eq.(3), and  using Eq.(1) with $M_{bh} = M_x = {M_0 \over x}$. 
Even though steady state  is not a good approximation for late stages of 
black hole evaporation, we will take this profile for our case. In our case 
it will not represent the entire temperature profile at a given time. 
Rather one can think of this as temperature of plasma shells which expand 
and cool. Assumption of constancy of luminosity may 
then be valid for these plasma shells. In this respect it is important to note 
that essentially the same profile is obtained if we take $ T = T_x$ effectively 
at a distance of order of the Schwarzchild radius  $r_x = {2 \over M_{pl}^2} 
{M_0 \over x}$ along with the condition $T^4 r^2 =$ constant (which essentially 
means conservation of energy-momentum of the expanding shells). 

Temperature will drop to the value $\eta$ at a distance $r_\eta$ which is 
obtained by setting $T(r_\eta) = \eta$ for $T(r)$ (obtained from Eq.(3)). 
String loops will be produced at this distance and will be dragged (and 
stretched) by expanding plasma.  The plasma flow should
stop at a distance of order $r_U$ where the temperature drops to the
ambient value $T_U$. $r_U$, therefore, sets an upper limit on the distance 
$R_{stretch}$ up to which strings can be carried out by plasma flow, and is
determined by setting $T(r_U) = T_U$. Using Eqs.(2),(3), we get, 
$r_U \simeq 0.04 x M_{pl}^{1/3} T_U^{-4/3} $. To determine
$R_{stretch}$ we need to find when string motion is dominated
by friction forces arising due to the scattering
of plasma particles from the string. The dominant contribution to this comes 
from Aharanov-Bohm scattering (of particles with appropriate fractional 
charges) for gauge strings \cite{frict}. The magnitude of the friction 
force per unit length on the cosmic string is then given by \cite{frict},

\begin{equation}
F_{frict} \sim \beta T^3 {v \over \sqrt{1 - v^2}} ~,
\end{equation} 

\noindent where $v$ is the string velocity through the plasma, $T$ is the 
plasma temperature, and $\beta$ is a numerical parameter related to
the number of relevant particle species \cite{frict}. In specific grand 
unified theory models \cite{fracq} it could be of order one,
and we will assume that to be the case here. (See, also, ref. \cite{evrt} 
for early discussions of scattering of particles from strings.) We mention 
here that stretching of string due to plasma flow has been discussed by 
Chudnovsky and Vilenkin in ref. \cite{curv} where they considered 
light superconducting cosmic strings getting stretched due to turbulent 
plasma flow in the galactic disc and even in stellar interiors. One  could
also consider formation and stretching of such strings around primordial
black holes. This will require considerations of magnetic fields in the
cosmic plasma etc.

  We consider a simple situation when a string encircling the black hole
is stretched symmetrically by radial flow of the plasma. (Order of 
magnitudes of our estimates should remain same for other geometries of 
strings, e.g. string ends may lie on the boundary of the symmetry
restored region around the black hole, as seen often in the liquid 
crystal case.) Such large strings should easily form during string
formation on the surface of the sphere with radius $r_\eta$ near the
black hole (as also seen in Fig.1 for the liquid crystal case).
String tries to collapse due to its tension. For a string
loop of radius $R$, one can estimate \cite{curv} the tension force to 
be of order $F_{tension} \sim {\mu \over R} ~=~ {\eta^2 \over R}$.

At the formation stage, near $r \simeq r_\eta$, the  string will be at 
rest in the local (expanding) plasma frame with $v = 0$ in Eq.(5), and 
hence $F_{frict} = 0$.  String loop will 
then start collapsing under its tension (in the local rest frame of the 
plasma). This will lead to a  non-zero $v$, and consequently a non-zero 
friction force via Eq.(5). String loop will expand in the rest frame of 
the black hole, due to drag of the plasma expanding with sound velocity, 
when $F_{frict}(v=1/\sqrt{3}) > F_{tension}$. This condition gives the 
upper limit for $R_{stretch}$ as,

\begin{equation}
R_{stretch} < 10^{-8} \times {x^3 M_{pl}^6 \over M_0^3 \eta^4} 
\equiv R_{max} ~.
\end{equation}

 As plasma flow stops beyond a distance of order $r_U$, we conclude that 
plasma flow can stretch string loops to sizes of order $R_{stretch}
\simeq min(R_{max}, r_U)$. We should also consider the effect of
black hole gravity on the string loops. Gravitational force per unit length
on the string due to the black hole of mass $M_{bh}$ can be roughly
estimated as ${G M_{bh} \mu \over R^2}$ where R is the separation of the 
string segment from the black hole. (We are taking string as a simple
gravitating system, as for a string loop. For special geometries, such
as for a straighter string the gravitational force will be different, and
will be typically smaller). We see that,
per unit length, this gravitational force becomes much less than the 
force $f_{tension} \sim {\mu \over R}$ due to string tension when
$R > G M_{bh}$. Thus for any distances much larger than the Schwarzchild
radius of the black hole, gravity of the black hole is subdominant
compared to string tension forces. We are considering the situation
when plasma drag forces completely dominate over the string tension forces, 
leading to stretching of string to large distances (where black hole gravity 
becomes even more negligible). Thus black hole gravity is not 
relevant for string dynamics  in our case. 

\begin{table}[h]
\begin{center}
\caption{Sample parameter values for percolation of strings}
\begin{tabular}{||ccccccc||}
\hline
f &$T_U$&$\eta$&$M_{0}/M_{pl}$&$M_x/M_{pl}$&$r_\eta$& $R_{stretch}$\\
 & GeV & GeV & & & GeV$^{-1}$ & $GeV^{-1}$\\
\hline
1&    $2 \times 10^7$ & $10^{10}$ &  $5 \times 10^6$ &    4000 & $10^{-7}$& 0.02 \\
$10^{-3}$ & $10^7$ &  $10^{10}$  & $8 \times 10^6$   &  550   & $6 \times 10^{-7}$&  0.7  \\
1.0 &  $10^5$  &  $10^8$ &  $10^8$  & 42000 & $8 \times 10^{-5}$&   87  \\
$10^{-3}$ & $10^4$ & $10^8$ &  $8 \times 10^8$  &  12000  & $3 \times 10^{-4}$& 30000 \\
$10^{-7}$ & $10^3$ & $10^8$ &  $4 \times 10^9$ & 1500  & $2 \times 10^{-3}$& $2 \times 10^7$ \\
1  & 100 &  $10^4$  & $2 \times 10^{10} $ &  $ 10^6 $ & 400 &$ 4 \times 10^6$ \\
$10^{-7}$ & 1.0  & $10^4$  & $4 \times 10^{11}$ & 33000  & $10^4$&  $10^{12}$ \\
1 &  1.0 &  100 &  $4 \times 10^{11}$ & $ 7 \times 10^6$  & $5 \times 10^5$ &  $ 5 \times 10^9$ \\
$ 10^{-7}$ &  1.0  &  100  & $ 4 \times 10^{11}$  & 33000 & $10^8$ & 
$10^{12}$ \\

\hline
\end{tabular}
\end{center}
\end{table}

If black hole number density in the universe is such that inter-black hole 
separation $d_{bh}$ is smaller than $R_{stretch}$ then loop intersections 
will be frequent. This will lead to intercommutation of strings with 
probability (almost) 1, implying percolation of strings, resulting in 
an infinite string network. By assuming that energy density of black holes
contributes only a fraction $f$ of the total energy density of
the universe, we get (with $g_* \simeq 100$),

\begin{equation}
d_{bh} \simeq {M_{pl}^{5/9} \over 8 f^{1/3} T_U^{14/9}} ~.
\end{equation}

 Percolation of loops (forming an infinite string network) will happen 
when $d_{bh} \le R_{stretch}$. With $R_{stretch} = r_U$ (when $r_U <
R_{max}$), this gives the minimum allowed value of $x$ as $x_{min} \simeq 
3 f^{-1/3} (M_{pl}/T_U)^{2/9}$ (consequently, maximum value of $M_x$). 
We will give results with $x = x_{min}$ 
(so that $d_{bh} = R_{stretch}$). We find that for string percolation, 
with $f = 1, 10^{-3}, 10^{-5}$, and $10^{-7}$, $\eta$ must be below 
about $2  \times 10^{14}$, $10^{12}, 2 \times 10^{10}$, and $5 \times 
10^8$ GeV respectively, while $T_U$ must be less than about $2 \times 
10^{13}$, $10^9, 3 \times 10^6$, and $5 \times 10^3$ GeV respectively.
(Note that $M_x$ should be large enough to create many strings of sizes 
$\sim R_{stretch}$. This is always true with our parameters.)
When $r_U > R_{max}$ then $R_{stretch} = R_{max}$. $\eta$ can be larger
in this case (e.g., up to $4 \times 10^{15}$ for $f = 1$). However, for
very large $\eta$, $M_x$ cannot be much bigger than $M_{pl}$. Use of
Eq.(1) may be suspect for such small black holes. We will quote results
for parameter values when $r_U < R_{max}$, and will not consider very
large values of $\eta$. 

 In Table 1 we have given several sets of values for various parameters 
for which string percolation happens. We see that $R_{stretch} >> r_\eta$,
meaning that the strings are dragged (via friction forces) and 
stretched by the plasma flow up to very large distances. 
We also see that the percolation of strings is not 
a rare occurrence, but happens for rather generic set of parameter values. 
For example, even with almost negligible fraction of energy density in 
black holes, $f = 10^{-7}$, string percolation is achieved.

   We conclude by emphasizing that our results present first possible
scenario where string loop generation by localized sources (heated
wire tip for the liquid crystal experiment, and primordial black holes for
the early universe) can lead to the formation of an extended string
network without an overall phase transition, and even when the ambient 
temperature is much below the string scale. During scaling regime, 
there may not be any difference in the properties of string networks 
formed via the mechanism discussed here, and those arising from the 
conventional mechanism during a phase transition. These results illustrate 
the tremendous potential of correspondence between phenomenon observed in 
condensed matter systems and those which could have occurred in the early 
universe.

\vskip .3in

 I would like to thank R. Rangarajan, S. Digal, S. Sengupta, R. Ray, 
B. Layek, and A.P. Mishra  for useful discussions and comments. I am
also grateful to the participants of the workshop "Laboratory Cosmology",
2006, at the  Lorentz Centre, Leiden, for useful discussions.

%%%%%%%%%%%%%%%%%%%

\vskip 1in
\centerline{\bf Figure Caption}
\vskip .3in

{\bf Fig.1:}

Production of string loops by heated wire tip embedded inside
NLC droplet. Strings are carried out by convection, and are stretched 
to large sizes

%\begin{figure}
%\epsfig{file=figlb.eps,height=125mm}
%\caption{Production of string loops by heated wire tip embedded inside
%a NLC droplet. Strings are carried out by convection, and are stretched 
%to large sizes}
%\label{fig:fig1}
%\end{figure}

\newpage

\end{document}